\documentclass[seceq]{ptptex}

\usepackage{amsmath}

\renewcommand{\theequation}{\thesection.\arabic{equation}}

\newcommand{\p}{\partial}
\newcommand{\Tr}{{\rm Tr}}
\newcommand{\Str}{{\rm Str}}

\title{
Fluctuation Analysis of the Non-Abelian 
Born-Infeld Action in Background Intersecting D-branes}

\author{Satoshi \textsc{Nagaoka}\footnote{{\tt
E-mail: nagaoka@hep1.c.u-tokyo.ac.jp}}}

\inst{{\it Institute of Physics, University of Tokyo, Tokyo 153-8902,
Japan}}

\recdate{October 1, 2003}

\abst{We carry out a fluctuation analysis of the non-abelian Born-Infeld 
action up to order $F^6$ 
in the presence of a background intersecting D-branes system. 
We compare the mass spectrum which is obtained in our analysis
to the spectrum derived from the worldsheet analysis.
The mass spectra completely agree with each other.
This result provides a strong support for the claim that the action 
we analyze here, which was proposed in hep-th/0208044 
by Koerber and Sevrin, is the correct low energy effective action
of string theory. }

\notypesetlogo

\begin{document}

\maketitle

\section{Introduction}

It is very important to consider D-brane dynamics in string theory.
For slowly varying fields,
the low energy dynamics of a D-brane is described by the Dirac-Born-Infeld 
action. For $N$ coincident D-branes, the gauge group becomes $U(N)$ and 
the leading low energy limit is described as the $U(N)$ super
Yang-Mills theory. \cite{wit} 
It is desirable to know the full non-abelian effective action 
for slowly varying fields. 
There are some reasons for the difficulty to describe an explicit 
form of the non-abelian Born-Infeld action.
The main problem in this regard is the ordering ambiguity.
The field strengths do not commute with each other, and 
the ordering of terms can not be determined without further information.
In addition, there is also ambiguity involved in expressing
the non-abelian Born-Infeld action 
in terms of either field strength or covariant derivatives.
Therefore the expansion under slowly varying fields becomes 
ambiguous. It was proposed that higher order corrections are 
given by the non-abelian generalization of the Born-Infeld action 
with a symmetrized trace. \cite{tse} 
There are some problems with this proposal at order $F^6$, 
that is, the fluctuation 
spectrum of this action in a constant magnetic background
differs from the spectrum 
obtained by the worldsheet analysis \cite{BDL}, as pointed out 
in Ref.~\citen{AkiWati}.
The mass spectrum was further examined in Ref.~\citen{DST},
and a new method to determine the 
non-abelian Born-Infeld action was developed in Ref.~\citen{FKS}. 
The authors of that work started from BPS solutions to the Yang-Mills 
action that define  
stable holomorphic bundles and considered deformation of the 
Yang-Mills action.
They required that stable holomorphic
bundles satisfy the equations of motion and 
constructed higher order terms of $F$ from lower order terms.
In the abelian case, they found that the deformation is given 
uniquely by the abelian Born-Infeld action. 
This method has been applied to construct the $F^5$ non-abelian effective
action \cite{F^5action}. This action  
has been examined by the fluctuation analysis
in a magnetic background, \cite{F^5fluc}
and agreement with the worldsheet analysis has been found.
This action was also confirmed through direct calculation 
of the open superstring 5-point amplitude \cite{MBM}.
$F^6$ non-abelian Born-Infeld terms were also proposed in
Ref.~\citen{KS} in the same way, but no tests have yet been carried out 
up to this order.

Recently, we considered the recombination in
non-supersymmetric intersecting D-branes system at one angle 
in the Yang-Mills Higgs theory. \cite{Recombi}\footnote{
The classical decay of unstable intersecting D-branes was also discussed in 
Ref.~\citen{Morosov}. }
In that system,
the tachyon mode localized at the intersection point which connects 
two D-strings condenses and that causes the recombination.
In Ref.~\citen{Recombi}, we obtained 
the correct mass of the tachyon mode in the limit of small intersection angle.
Therefore, by comparing the mass spectrum obtained in this analysis employing
the action with higher $F$ corrections to
the result obtained from the worldsheet analysis,
we can examine the non-abelian Born-Infeld action.

In this paper, we examine the non-abelian Born-Infeld action
at order $F^6$ proposed in Ref.~\citen{KS} by carrying out 
fluctuation analysis in the background intersecting brane system.
This action has terms with covariant derivatives that cannot be eliminated
by the field redefinition and necessarily remain.
Therefore, the action at this order becomes much more 
complicated than that of lower orders, $F^4$ or even $F^5$. 
Therefore, analysis up to this order is desirable in order to claim that 
the method proposed in Ref.~\citen{FKS} is appropriate.
In \S 2, we present the mass spectrum which was obtained 
by the worldsheet analysis and also summarize the fluctuation 
analysis for the $F^2$ term and the $F^4$ terms in this system.
In \S 3, we calculate the mass spectrum of the fluctuations for the 
action with the $F^5$ terms.
In \S 4, we calculate the mass spectrum for the action with 
the $F^6$ terms.
Most of the calculations presented here were carried out using Mathematica.
Section 5 is devoted to discussion.
A detailed calculation of the $F^6$ action is given in Appendix A.

\section{Fluctuation analysis up to ${\cal O}(F^4)$ }

Before considering the fluctuation analysis, 
we present the mass spectrum of a string stretching between 
two D-strings intersecting at angle $\theta$ that is obtained 
by the worldsheet analysis. \cite{BDL} This is given by
\begin{align} \label{wsspec}
m_n^2=\left(n-\frac{1}{2}\right)\frac{\theta}{\pi\alpha'}\ ,
\end{align}
where $n$ is $0,1,\cdots$.
In this section, we briefly explain the fluctuation analysis 
for the Yang-Mills action and for the non-abelian Born-Infeld action 
up to ${\cal O}(F^4)$. These were discussed in Ref.~\citen{Recombi} in detail.

\subsection{Fluctuation analysis in the intersecting brane system}

Let us consider the low energy superstring effective action 
without Born-Infeld corrections, i.e., the Yang-Mills action. 
The (2+1) dimensional SU(2) Yang-Mills action is given by
\begin{align}
S=-\frac{1}{4} {\rm Tr } \int\!dtd^2x  F_{i_1i_2}^2 \ .
\end{align}
The worldvolume coordinates are taken as 0,1 and 2. 
By taking the T-duality along the $2$ direction, 
we obtain the Lagrangian with transverse scalar fields in the bulk
as\footnote{We have absorbed the factor $2\pi\alpha'$ of the second terms 
through a field redefinition of $\Phi.$}
\begin{align} 
L_0=-{\rm Tr }  \left(\frac{1}{4} F_{\mu\nu}^2 
+\frac{1}{4}(D_\mu \Phi)^2\right) \ ,
\end{align}
where $\mu,\nu=0,1$, and 
the covariant derivative is defined as 
\begin{align}
D_\mu \Phi =\p_\mu \Phi -i[A_\mu,\Phi] \ .
\end{align}
This Lagrangian constitutes the low energy description of two D-strings.

Now, we consider the intersecting D-strings background and discuss 
fluctuation around this background.
The reason that we treat the Yang-Mills Higgs Lagrangian which emerges 
through dimensional reduction from the original Yang-Mills Lagrangian
by taking the T-duality is because this results in calculation that is
simpler than that in the case of the original Yang-Mills Lagrangian.
The classical solution that represents the intersecting D-strings 
is given by
\begin{align} \label{sol}
\Phi=qx\sigma^3 , \quad A_\mu=0 \ .
\end{align}
This solution represents D-strings intersecting at an angle $\theta
\equiv 2 \tan^{-1} (2 \pi\alpha' q)$.

The fluctuations we turn on here are expressed as 
\begin{align} \label{fluc}
\Phi=qx \sigma^3+ f(x_\mu) \sigma^1, \quad
A_1= g(x_\mu) \sigma^2 \ .
\end{align}
There is another combination of off-diagonal fluctuations 
for which $\Phi \propto \sigma^2$ and $A_1 \propto \sigma^1$, but
it is sufficient to consider only the fluctuations (\ref{fluc}),
because other 
fluctuations decouple from these ones at the quadratic level.\footnote{
These two combinations correspond to the modes connecting 
two branes with opposite directions.
In the analysis of the non-abelian Born-Infeld action which is discussed
later, we assume that the combination of the fluctuations $f$ and $g$ 
decouple from other fluctuations at the quadratic level.}
The Lagrangian quadratic in the fluctuations is given by
\begin{align} \label{quad} 
\tilde{L}_0=\left[(\p_t f)^2 +  (\p_t g)^2 \right]
-\left[(\p_x f)^2+4 q f g- 4 q (\p_x f) g x +4 q^2 g^2 x^2 \right]\ .
\end{align}
The equations of motion for this Lagrangian are obtained as
\begin{align}
\hat{O}_0
\left(
\begin{array}{c}
f\\ g
\end{array}\right)=0 \ ,
\end{align} 
where the differential operator $\hat{O}_0$ is written 
\begin{align}
\hat{O}_0=
-2\left(
\begin{array}{cc}
- \p_x^2 +\p_t^2& 4q+2qx \p \\ 2q-2qx \p& 4q^2x^2 +\p_t^2
\end{array}\right) \ .
\end{align}
Expanding the fluctuations in the mass eigen functions,
\begin{align} \notag
f(x,t)=\Sigma u_n (x) C_n (t) \ , \\
g(x,t)=\Sigma v_n (x) C_n (t) \ ,
\end{align}
the equation of motion for the fluctuation fields of the spatial part 
is obtained as
\begin{align} \label{diffeq}
\left(
\begin{array}{cc}
- \p_x^2& 4q+2qx \p_x \\ 2q-2qx \p_x& 4q^2x^2
\end{array}\right)
\left(
\begin{array}{c}
u_n \\ v_n 
\end{array}\right)
=m_n^2\left(
\begin{array}{c}
u_n\\ v_n
\end{array}\right)  \ ,
\end{align} 
and the decomposed field $C_n (t)$ satisfies the free field equation 
\begin{align}
(\p_t^2+m_n^2)C_n(t)=0 \ ,
\end{align}
with the mass eigenvalues $m_n$.
The mass squared $m^2_n$ can be obtained by solving (\ref{diffeq}),
and the result is 
\begin{align} \label{massf^2}
m^2_n=\left(n-\frac{1}{2}\right) 4q 
=\left(n-\frac{1}{2}\right)\frac{2 \tan (\theta/2)}{\pi\alpha'} \ .
\end{align}
The corresponding normalizable eigen functions are obtained as
\begin{align}
\notag
u_n (x)= e^{-qx^2}\sum_{j=0,2,\cdots}^n (-1)^{\frac{j}{2}} 
\frac{2^{j}}{j!}
\frac{n(n-2)\cdots (n-j+2)}{2n-1} (2n\!-\!j\!-\!1) 
\left(x \sqrt{q} \right)^j \ , \\
v_n(x) =- e^{-qx^2} \sum_{j=0,2,\cdots}^n (-1)^{\frac{j}{2}} 
\frac{2^{j}}{j!}
\frac{n(n-2)\cdots (n-j+2)}{2n-1} (j-1) 
\left( x\sqrt{q}\right)^j \ ,
\label{geneven}
\end{align}
for $n=0,2,\cdots$, and
\begin{align}\notag
u_n(x)=&
 e^{-qx^2}
\Big(\sqrt{q} x  \\ \notag 
&+\sum_{j=3,\cdots}^n (-1)^{\frac{(j-1)}{2}} 
\frac{2^{(j-1)}}{j!}
\left(n-\frac{j+1}{2}\right) (n-3) \cdots (n-j+2)
\left(x \sqrt{q} \right)^j \Big) \ , \\
v_n(x)=&
- e^{-qx^2}\sum_{j=1,3,\cdots}^n (-1)^{\frac{(j-1)}{2}}
 \frac{2^{(j-1)}}{j!}
\left(\frac{j-1}{2}\right) (n-3) \cdots (n-j+2)
\left(x \sqrt{q} \right)^j \ , 
\label{genodd}
\end{align}
for $n=3,5,\cdots$. 
Considering the lowest mode $n=0$,
the mass squared is negative and given by
\begin{align} \label{ymspec}
m^2_0= -2q = -\frac{\tan (\theta /2)}{\pi \alpha'} \ .
\end{align}
The corresponding eigen functions are Gaussian and given by
\begin{align} \notag
u_0(x)=e^{-q x^2}=\exp\left[{-\frac{\tan (\theta /2)}{2 \pi \alpha'}} x^2
\right] \ , \\
v_0(x)=-e^{-q x^2}=-\exp\left[{-\frac{\tan (\theta /2)}{2 \pi \alpha'}} x^2
\right] \ . \label{gaussian}
\end{align}
The mass tower (\ref{massf^2}) is expanded in $\theta$ as
\begin{align} \label{massexp}
m_n^2=\left(n-\frac{1}{2}\right) 
\frac{1}{2\pi\alpha'}\left(\theta +{\cal O} (\theta)^3\right) \ .
\end{align}
Therefore, comparing (\ref{massexp}) to (\ref{wsspec}), we conclude
that the string mass spectrum is obtained correctly 
at ${\cal O}(\theta)$ in the Yang-Mills analysis.

\subsection{$F^4$ corrections}

We consider the non-abelian Born-Infeld corrections at order $F^4$
in this subsection.
We can obtain the $F^4$ terms by expanding the Lagrangian
\begin{align}
L={\rm Str} \sqrt{- \det (\eta_{i_1i_2} +2 \pi\alpha' F_{i_1i_2})} \ ,
\end{align}
and the result is 
\begin{align}
L_2={\rm Str} (2\pi\alpha')^2
\left(\frac{1}{8} F_{i_1i_2} F_{i_2i_3} F_{i_3i_4} F_{i_4i_1}
-\frac{1}{32} F_{i_1i_2} F_{i_2i_1} F_{i_3i_4} F_{i_4i_3}\right) \ ,
\end{align}
where $\rm Str$ denotes the symmetrized trace.

For the Lagrangian up to this order, 
we take the T-duality along the 2 direction and  
carry out the fluctuation analysis,
which is the same as that given in the previous subsection.
The classical solution is identical to
(\ref{sol}), because we consider the diagonalized solution.
The Lagrangian quadratic in the fluctuations is obtained as
\begin{align} \notag
\tilde{L}=\tilde{L}_0+\tilde{L}_2
=&\left( 1-\frac{1}{6} (2\pi\alpha' q)^2 \right) \left[(\p_t f)^2 
+  (\p_t g)^2 \right]\\
&-\left(1-\frac{1}{2}(2\pi\alpha'q)^2 \right)
\left[(\p_x f)^2+4qfg-
4 q (\p_x f) g x +4 q^2 g^2 x^2 \right] \ , \label{f^4quad}
\end{align}
and the corresponding equations of motion for $f$ and $g$ are 
\begin{align}
\left(\hat{O}_0+\hat{O}_2\right)
\left(
\begin{array}{c}
f\\ g
\end{array}\right)=0 \ ,
\end{align} 
where the differential operator derived 
from the $F^4$ terms is written
\begin{align}
\hat{O}_2=
(2\pi\alpha' q)^2 \left(
\begin{array}{cc}
- \p_x^2 & 4q+2qx \p \\ 2q-2qx \p& 4q^2x^2 
\end{array}\right)
+\frac{1}{3}(2\pi\alpha' q)^2
\left(
\begin{array}{cc}
 \p_t^2 & 0 \\ 0& \p_t^2
\end{array}\right) \ .
\end{align}

When we compare this fluctuation Lagrangian (\ref{f^4quad}) 
to the Yang-Mills result (\ref{quad}), 
we find that the form of the mass term is unchanged, 
except for an overall factor.
Therefore, the eigen functions are unchanged, 
and the modification emerges in the spacing of the mass tower as follows:
\begin{align}
m_n^2 = (2n-1)2q\left( 1-\frac13 (2 \pi\alpha' q)^2 \right) 
= \left(n-\frac12\right)\frac{\theta}{\pi\alpha'} 
+ {\cal O} (\theta^5) \ .
\end{align}
Therefore, we obtain the correct ${\cal O } (\theta^3)$ corrections.

\section{$F^5$ corrections}

The non-abelian Born-Infeld Lagrangian at order $F^5$ 
was proposed in Ref.~\citen{KS} 
as\footnote{The overall sign is different from that in Ref.~\citen{KS}. 
This results from the difference in the choice of the basis.
We choose a hermitian basis here, and they chose 
an anti-hermitian basis.}
\begin{align}\label{F^5ac}
L_3=-{\rm Str}(2\pi\alpha')^3 \frac{\zeta (3)}{2 \pi^3}
\Tr ([D_{i_3},D_{i_2}] D_{i_4} F_{i_5 i_1} 
D_{i_5}[D_{i_4},D_{i_3}] F_{i_1,i_2}) \ . 
\end{align}
As in the analysis given in the previous section,
we take the T-duality and consider off-diagonal fluctuations 
around the background intersecting brane system at angle $\theta$. 
A straightforward calculation leads to the 
Lagrangian quadratic in the fluctuations as
\begin{align}
\tilde{L}=\tilde{L}_0+\tilde{L}_2+\tilde{L}_3 \ ,
\end{align}
and the contribution from the Lagrangian of the $F^5$ term is given by
\begin{align} \notag
\tilde{L}_3=&-(2\pi\alpha')^3 \frac{\zeta (3)}{2 \pi^3} q^2 
\\ &\cdot
\left(
-32 \dot{f}^2 q^2 x^2 -8\ddot{f}^2+8(\dot{f}^{'2}-8 \dot{f}' \dot{g} q
 x+4 \dot{g}^2 q^2 x^2)-8 \ddot{g}^2-8 \dot{g}^{'2}
\right) \ ,
\end{align}
where $\dot{f}(x,t)$ and $f'(x,t)$ denote the derivatives 
of $f(x,t)$ with respect to $t$ and $x$ respectively.
The equations of motion for this Lagrangian are given by
\begin{align} \label{F^5eom}
\left(\hat{O}_0+\hat{O}_2+\hat{O}_3\right)
\left(
\begin{array}{c}
f\\ g
\end{array}\right)=0 \ ,
\end{align} 
where the $F^5$ part of the differential operator is written 
\begin{align} \label{ohat3}
\hat{O}_3=
-16(2\pi\alpha')^3 \frac{\zeta (3)}{2 \pi^3}
q^2 \p_t^2 \left(
\begin{array}{cc}
4 q^2 x^2 -\p_t^2 + \p_x^2
&-4 q (1+x\p_x)
\\ 
4 q x \p_x
& - \p_t^2 -\p_x^2 -4 q^2 x^2 
\end{array}\right) \ .
\end{align}
To find the solution of (\ref{F^5eom}), 
let us consider the form of the solution as
\begin{align} \notag
f_0^{\text{sol}}(x,t)&=e^{-qx^2}e^{im_0^{\text{sol}}t} \\
g_0^{\text{sol}}(x,t)&=-e^{-qx^2}e^{im_0^{\text{sol}}t} 
\label{f50ans} \\
(m_0^{\text{sol}})^2&=-2q
\left(1-\frac{1}{3}(2\pi\alpha'q)^2
+a_1(2\pi\alpha'q)^3+b_1(2\pi\alpha'q)^4\right) +{\cal O} (q^6) \ , \notag
\end{align}
where we ignore the normalization constant.
We have taken the spatial dependence of the solution to be 
equivalent to the tachyon mode 
of the leading order result (\ref{gaussian}).
Therefore, this solution should correspond to the tachyonic lowest mode.
This solution satisfies the equations of motion (\ref{F^5eom}) when $a_1=b_1=0$
up to ${\cal O} (q^5)$. 
However, what we want to obtain is the mass for the equations of motion 
(\ref{F^5eom}), and therefore we must diagonalize the $\p_t$ operators in 
(\ref{ohat3}) so that they are proportional to $\p_t^2 \cdot {\bf \hat{1}}$.
We carry out this procedure through the field redefinition given by\footnote{
In (\ref{f^5red}), ${\cal O} (q^6)$ 
is equivalent to ${\cal O}\left((2\pi\alpha')^6 \right)$.}
\begin{align} 
\left(
\begin{array}{c}
\tilde{f}\\ \tilde{g}
\end{array}\right)
= \left(
\hat{{\bf 1}}
+4(2\pi\alpha')^3 \frac{\zeta (3)}{2 \pi^3} q^2
\left(
\begin{array}{cc}
-\p_t^2+4 q^2 x^2
&-4qx\p_x
\\ 0&-\p_t^2-\p_x^2
\end{array}\right)+{\cal O} (q^6)
\right)
\left(
\begin{array}{c}
f\\ g
\end{array}\right) \ .
\label{f^5red}
\end{align}
Under this transformation, the time derivative operators 
in the equations of motion (\ref{F^5eom}) become
space derivative operators, and the redefined fields
$\tilde{f}$ and $\tilde{g}$ obey the deformed equations of motion
\begin{align}  \label{f^5redeom}
\left(\hat{O}_0+\hat{O}_2+{\hat{O}_3}'\right)
\left(
\begin{array}{c}
\tilde{f}\\ \tilde{g}
\end{array}\right)=0 \ ,
\end{align}
where
${\hat{O}_3}'$ is the differential operator deformed through 
the field redefinition and is given by
\begin{align}\notag
{\hat{O}_3}'&=
-16(2\pi\alpha')^3 \frac{\zeta (3)}{2 \pi^3}q^2 \\
&\left(
\begin{array}{cc}
4 q^2 x^2 \p_x^2+8 q^2 x \p_x +4q^2
&-4 q^3 x^3 \p_x -8 q^3 x^2 -qx\p_x^3-2 q\p_x^2
\\4q^3 x^3 \p_x +4 q^3 x^2 + qx\p_x^3+ q \p_x^2
&-4 q^2 x^2 \p_x^2-8 q^2 x \p_x-4q^2
\end{array}\right) 
+{\cal O}(q^7)  . 
\end{align}
The time dependence of the solution (\ref{f50ans}) remains 
unchanged under the field redefinition (\ref{f^5red}),
and therefore the redefined solution is easily obtained by substituting 
(\ref{f50ans}) into (\ref{f^5red}) as
\begin{align} \notag
\tilde{f}_0^{\text{sol}}&=\left(1-8(2\pi\alpha' q)^3
\frac{\zeta (3)}{2 \pi^3}
(1+ 2 q x^2)+{\cal O} (q^6)
\right)e^{-qx^2}e^{im_0^{\text{sol}}t} \ , \\
\tilde{g}_0^{\text{sol}}&=-\left(1-16(2\pi\alpha' q)^3
\frac{\zeta (3)}{2 \pi^3}
qx^2+{\cal O}(q^6)
\right)e^{-qx^2}e^{im_0^{\text{sol}}t} \ .
\end{align}
We obtain $a_1=b_1=0$ again, which are now the corrections 
for the mass eigenvalue\footnote{
There is some ambiguity in 
the field redefinition that diagonalizes the time derivative operator 
in (\ref{f^5red}).
This is due to the fact that we can choose 
whether to absorb the $\dot{f}\dot{g}$ term through the
redefinition of $f$ or $g$.
When we change the form of the field redefinition, 
the redefined operator also changes,
and there is no ambiguity in
the corresponding mass eigenvalue up to this order.}
by solving the equations of motion 
(\ref{f^5redeom}) up to ${\cal O}(q^5)$.
Therefore, the action at order $F^5$ gives no contribution to the 
mass of the lowest tachyon mode at both ${\cal O}(\theta^4)$ 
and ${\cal O} (\theta^5)$.

Next, we consider the $F^5$ correction for the massive mode.
The ansatz we consider here is 
\begin{align} \notag
f_n^{\text{sol}}(x,t)&=u_n^{\text{sol}}(x)C_n^{\text{sol}}(t)\ ,\\ \notag
g_n^{\text{sol}}(x,t)&=v_n^{\text{sol}}(x)C_n^{\text{sol}}(t)\ ,\\ \notag
C_n^{\text{sol}}(t)&=e^{im_n^{\text{sol}}t}\ ,\\ 
(m_n^{\text{sol}})^2&=(4n-2)q
\left(1-\frac{1}{3}(2\pi\alpha'q)^2
+a_{n1}'(2\pi\alpha'q)^3+b_{n1}'(2\pi\alpha'q)^4\right)+{\cal O}(q^6) \ ,
\label{f5ans}
\end{align}
where $u_n^{\text{sol}}(x)$ and $v_n^{\text{sol}}(x)$ are the solutions 
of the Yang-Mills analysis given in (\ref{geneven}) and (\ref{genodd}). 
This solution satisfies the equations of motion (\ref{F^5eom})
when $a_{n1}'=b_{n1}'=0$ up to ${\cal O}(q^5)$.
The field redefinition (\ref{f^5red}) leads to the equations of motion
(\ref{f^5redeom}), and the redefined solution for the massive mode 
is represented by
\begin{align} \notag
\tilde{f}_n^{\text{sol}}&=
\left(u_n^{\text{sol}}+\frac{1}{2}(2\pi\alpha'q)^2
\left(-16 q u_n^{\text{sol}}+32 q^2 x^2 u_n^{\text{sol}}
-32qx\p_x v_n^{\text{sol}}
\right)
\right)e^{im_n^{\text{sol}}t}
\\ \notag
&=e^{im_n^{\text{sol}}t}
\Biggm[u_n^{\text{sol}}(x)-8(2\pi\alpha' q)^3  \frac{\zeta (3)}{2\pi^3}
e^{-qx^2} \\ \notag
&
\Bigg(
1+\sum_{j=2,4,\cdots}^{n+2}(-1)^{\frac{j}{2}}\frac{2^j}{j!}
\frac{n(n-2)\cdots(n-j+4)}{2(2n-1)} \\ \notag 
& \hspace*{3cm}
(-4-j+j^3+6n-4jn-2j^2n+4n^2)(x\sqrt{q})^j
\Bigg)
\Biggm] , \\
\notag
\tilde{g}_n^{\text{sol}}&=
\left(v_n^{\text{sol}}+\frac{1}{2}(2\pi\alpha'q)^2
\left(
-16 q v_n^{\text{sol}}-8 \p^2_x v_n^{\text{sol}}
\right)
\right)e^{im_n^{\text{sol}}t}
\\ \notag
&=e^{im_n^{\text{sol}}t}
\Biggm[v_n^{\text{sol}}(x)+4(2\pi\alpha' q)^3 \frac{\zeta(3)}{2\pi^3}
 e^{-qx^2} \\ \notag
& \biggm(-\frac{4n}{2n-1}
+\sum_{j=2,4,\cdots}^{n+2}(-1)^{\frac{j}{2}}
\frac{2^j}{j!}
\frac{n(n-2)\cdots(n-j+4)}{2n-1}   \\
& \hspace*{1cm}
(13j-4j^2-j^3-8n+4jn+4j^2n-4n^2-4jn^2)(x\sqrt{q})^j
\biggm)
\Biggm] ,
\end{align}
for $n=0,2,\cdots$, and
\begin{align}
 \notag
\tilde{f}_n^{\text{sol}}
&=e^{im_n^{\text{sol}}t}
\Biggm[u_n^{\text{sol}}(x)+8(2\pi\alpha' q)^3  \frac{\zeta(3)}{2\pi^3}
e^{-qx^2} \\ \notag
& \biggm(
-x\sqrt{q}+\frac{2(n-5)}{3}(x\sqrt{q})^3
+\sum_{j=5,7,\cdots}^{n+2}(-1)^{\frac{j-1}{2}}\frac{2^j}{j!}
\frac{(n-3)(n-5)\cdots(n-j+4)}{8}  \\
&\hspace*{4.5cm} \notag
(4+j-j^3-6n+4jn+2j^2n-4n^2
)(x\sqrt{q})^j
\biggm)
\Biggm] ,\\ \notag
\tilde{g}_n^{\text{sol}}
&=e^{im_n^{\text{sol}}t}
\Biggm[v_n^{\text{sol}}(x)+4(2\pi\alpha' q)^3  \frac{\zeta(3)}{2\pi^3}
e^{-qx^2} \\ \notag
& \biggm(-4 x\sqrt{q}+\frac{8(2n-3)}{3}(x\sqrt{q})^3
-\sum_{j=5,7,\cdots}^{n+2}(-1)^{\frac{j-1}{2}}
\frac{2^j}{j!}
\frac{(n-3)(n-5)\cdots(n-j+4)}{4}  \\
& \hspace*{1cm}
(3j-20j^2+9j^3+8n+4jn-12j^2n+4n^2+4jn^2)(x\sqrt{q})^j
\biggm)
\Biggm],
\end{align}
for $n=3,5,\cdots$.
These solutions satisfy the redefined equations of motion
(\ref{f^5redeom}) if $a_{n1}'=b_{n1}'=0$.
Therefore, we conclude that the $F^5 $ action (\ref{F^5ac}) does not
contribute to the mass spectrum up to ${\cal O}(q^5)$ which agrees
with the mass spectrum obtained by the worldsheet analysis.

\section{$F^6$ corrections}

The non-abelian Born-Infeld Lagrangian of $F^6$ terms 
proposed in Ref.~\citen{KS} is written 
\begin{align}
L_4=&L_{4,1}+L_{4,2}+L_{4,3} \ ,
\end{align}
where $L_{4,1}$ is given by
\begin{align}
L_{4,1}=
&(2\pi\alpha')^4
{\rm Str} 
\sum_{n=1}^3 
L_{41n} \ , \notag \\ 
L_{411}&=
\frac{1}{12} F_{i_1i_2} F_{i_2 i_3} F_{i_3i_4} 
F_{i_4i_5} F_{i_5i_6} F_{i_6i_1} \ , \quad
L_{412}=
-\frac{1}{32} F_{i_1i_2} F_{i_2i_3} F_{i_3i_4} F_{i_4i_1}
F_{i_5i_6} F_{i_6i_5} \ , \notag \\ 
L_{413}&=\frac{1}{384} F_{i_1i_2} F_{i_2i_1} F_{i_3i_4} F_{i_4i_3} 
F_{i_5i_6} F_{i_6i_5} \ ,
\end{align}
$L_{4,2}$ is given by
\footnote{Again,
the signs of some terms here are different from those in Ref.~\citen{KS}, 
as in the case of the $F^5$ action. 
This also results from the difference in the choice of the basis.
Apart from this, there is a typo in Ref. \citen{KS} that 
the overall sign of the $F^6$ terms flips.
This typo is corrected in Ref.~\citen{SW}.}
\begin{align} \notag
L_{4,2}=
&-(2\pi\alpha')^4
{\rm Str} 
\sum_{n=1}^8 L_{42n}  \\ 
L_{421}&=
-\frac{1}{24} F_{i_1i_2} D_{i_1} D_{i_6} D_{i_5} 
F_{i_2i_3} D_{i_6} F_{i_3i_4} F_{i_4i_5} \ , \quad \notag \\
L_{422}&=
-\frac{1}{48}
F_{i_1i_2} D_{i_5}D_{i_6} F_{i_2i_3} D_{i_6} D_{i_1} F_{i_3i_4} F_{i_4i_5} 
 \ , \notag \\
L_{423}&=
\frac{1}{24}F_{i_1i_2} [D_{i_6},D_{i_1}] D_{i_5} F_{i_2,i_3}
F_{i_3,i_4}D_{i_4}F_{i_5i_6}\ ,\quad \notag \\
L_{424}&=
\frac{1}{16}
D_{i_4}D_{i_5}F_{i_1i_2}F_{i_2i_3}[D_{i_6},D_{i_1}]F_{i_3i_4}F_{i_5i_6} 
 \notag \ , \\
L_{425}&=
\frac{1}{24}
D_{i_6}[D_{i_4},D_{i_5}]F_{i_1i_2}F_{i_2i_3}D_{i_1}F_{i_3i_4}F_{i_5i_6}
\ ,\quad \notag \\
L_{426}&=
\frac{1}{24} 
D_{i_6}D_{i_5}F_{i_1i_2}[D_{i_6},D_{i_1}]F_{i_2i_3}F_{i_3i_4}F_{i_4i_5}
\notag \ , \\
L_{427}&=
\frac{1}{24}
[D_{i_6},D_{i_1}]D_{i_3}D_{i_4}F_{i_1i_2}F_{i_2i_3}F_{i_4i_5}F_{i_5i_6}
\ ,\quad \notag \\
L_{428}&=
\frac{1}{48}
[D_{i_6},D_{i_4}]F_{i_1i_2}F_{i_2i_3}[D_{i_3},D_{i_1}]F_{i_4i_5}F_{i_5i_6} 
\ , \notag
\end{align}
and $L_{4,3}$ is given by
\begin{align} \notag
L_{4,3}=
(2\pi\alpha')^4
&{\rm Str} 
\sum_{n=1}^8 L_{44n}  \ , \\ \notag
L_{441}&=
\frac{1}{1440}
D_{i_6}[D_{i_4},D_{i_2}]D_{i_5}D_{i_5} [D_{i_1},D_{i_3}]D_{i_6}
F_{i_1i_2}F_{i_3i_4}\ , \\ \notag 
L_{442}&=\frac{1}{360}
D_{i_2}D_{i_6}[D_{i_4},D_{i_1}][D_{i_5},[D_{i_6},D_{i_3}]]D_{i_5}F_{i_1i_2}
F_{i_3i_4} \ , \\ \notag 
L_{443}&=\frac{1}{720}
D_{i_2}[D_{i_6},D_{i_4}][D_{i_6},D_{i_1}]D_{i_5}[D_{i_5},D_{i_3}]
F_{i_1i_2}F_{i_3i_4}\ , \\ \notag 
L_{444}&=\frac{1}{240}
D_{i_2}[D_{i_6},D_{i_4}]D_{i_5}[D_{i_6},D_{i_1}][D_{i_5},D_{i_3}]
F_{i_1i_2}F_{i_3i_4}\ ,  \notag 
\end{align}
\begin{align} \notag
L_{445}&=\frac{1}{360}
D_{i_6}D_{i_5}[D_{i_6},D_{i_4}][D_{i_5},D_{i_1}][D_{i_4},D_{i_3}]
F_{i_1i_2}F_{i_2i_3}\ , \\ \notag 
L_{446}&=\frac{1}{360}
D_{i_6}D_{i_5}[D_{i_4},D_{i_2}][D_{i_6},D_{i_1}][D_{i_5},D_{i_3}]
F_{i_1i_2}F_{i_3i_4} \ , \\ \notag 
L_{447}&=\frac{1}{360}
 D_{i_6}[D_{i_5},D_{i_4}][D_{i_3},D_{i_2}][D_{i_5},[D_{i_6},D_{i_1}]]
F_{i_1i_2}F_{i_3i_4}\ , \\
L_{448}&=\frac{1}{720}
[D_{i_6},D_{i_1}][D_{i_2},D_{i_6}][D_{i_5},D_{i_4}][D_{i_5},D_{i_3}]
F_{i_1i_2}F_{i_3i_4} \ .
\end{align}
$L_{4,1}$ is the symmetrized trace part of the non-abelian Born-Infeld 
action, and $L_{4,2}$ comprises the terms with four covariant derivatives 
and $L_{4,3}$ with eight covariant derivatives.

After taking the T-duality, we consider off-diagonal fluctuations 
around the background intersecting D-branes at an angle $\theta$,
as in the analysis given in the previous section.
A straightforward but lengthy calculation leads to 
the Lagrangian quadratic in the fluctuations as
\begin{align}
\tilde{L}=\tilde{L}_0+\tilde{L}_2+\tilde{L}_3+\tilde{L}_{4,1}+\tilde{L}_{4,2}
+\tilde{L}_{4,3} \ ,
\end{align}
and the contributions from the $F^6$ terms 
are given by
\begin{align} \label{L4-1}
\tilde{L}_{4,1}&=(2\pi\alpha'q)^4\left(
\frac{3}{40}(\dot{f}^2+\dot{g}^2)
-\frac{3}{8}(4 fgq+(f'-2 gq x)^2) 
\right) \ , \\
\tilde{L}_{4,2}+\tilde{L}_{4,3}&=
-(2\pi\alpha'q)^4 \biggm(
\frac{1}{45}\left(4 fgq+(f'-2 gq x)^2\right) 
\notag \\ \notag
&-\frac{1}{45}x^2(\dot{f}'-2q\dot{g}x)^2 
-\frac{1}{180 q^2}\left(4\dot{f}'q\dot{g}'
+(\dot{f}''-2q\dot{g}'x)^2\right)
\\ \notag
 &+\frac{1}{45}x^2(f''-2qg'x-4qg)^2+\frac{2}{45}x^2q^2(f-xf'+2qgx^2)^2 \\
&+\frac{1}{360q^2}(f'''-2qg''x-6qg')^2
+\frac{1}{360q^2}(\ddot{f}'-2 \ddot{g}qx)^2-\frac{1}{18q}\ddot{f}\ddot{g}
\biggm) \ . \label{L4-23}
\end{align}
We present this calculation term by term (19 terms) in detail in Appendix A.
The equations of motion for the total Lagrangian quadratic in the 
fluctuations are obtained as
\begin{align} \label{F^6eom}
\left(\hat{O}_0+\hat{O}_2+\hat{O}_3+\hat{O}_{4,1}+\hat{O}_{4,2}+\hat{O}_{4,3}
\right)
\left(
\begin{array}{c}
f\\ g
\end{array}\right)=0 \ ,
\end{align} 
where $\hat{O}_{4,1}$ and $\hat{O}_{4,2}+\hat{O}_{4,3}$ are given by
\begin{align}
\hat{O}_{4,1}=-\frac{3}{4}(2\pi\alpha'q)^4
\left(
\begin{array}{cc}
- \p_x^2 & 4q+2qx \p \\ 2q-2qx \p& 4q^2x^2 
\end{array}\right)
-\frac{3}{20}(2\pi\alpha' q)^4
\left(
\begin{array}{cc}
 \p_t^2 & 0 \\ 0& \p_t^2
\end{array}\right) \ ,
\end{align}
and
\begin{align} \notag
\hat{O}_{4,2}+\hat{O}_{4,3}=
&\frac{(2\pi\alpha' q)^4}{180} \\
\times  
\Biggm[
 &
16 q^2 x^2\left(
\begin{array}{cc}
-4 +4  x \p_x +x^2
 \p_x^2
&-12qx^2-2qx^3\p_x \\ 
-2qx^2+2qx^3\p_x
& -4q^2x^4
\end{array}\right) \notag \\
&+8\left(
\begin{array}{cc}
-  \p_x^2-4x\p_x^3-x^2\p_x^4
&2 q(2+13x\p_x+8x^2\p_x^2+x^3\p_x^3 ) \\ 
2q(-1+x\p_x-x^2\p_x^2-x^3\p_x^3)
& q^2(4x^2+16x^3\p_x +4x^4\p_x^2)
\end{array}\right) \notag \\
&+\frac{1}{q^2}\left(
\begin{array}{cc}
\p_x^6
& -12q \p_x^4 -2qx\p_x^5 \\ 
-2q\p_x^4+2 qx\p_x^5
& 4q^2(4\p_x^2
-4x\p_x^3-x^2\p_x^4)
\end{array}\right)\notag \\
+\frac{1}{q^2}\p_t^2&\left(
\begin{array}{cc}
16q^2 x\p_x +8q^2x^2\p_x^2
-2 \p_x^4
&-48q^3x^2-16q^3x^3\p_x+12q\p_x^2+4qx\p_x^3\\ 
16q^3x^3\p_x-4qx\p_x^3 &
-32q^4x^4+16q^2x\p_x+8q^2x^2\p_x^2
\end{array}\right) \notag \\
&+\frac{1}{q^2}\p_t^4\left(
\begin{array}{cc}
\p_x^2
&8q-2qx\p_x \\ 
10q+2qx\p_x & -4q^2x^2
\end{array}\right)
 \Biggm]
 \ . \label{O4-23}
\end{align}
The first term in (\ref{O4-23}) is derived from the fifth term in
the Lagrangian (\ref{L4-23}), the second term from the first and
fourth terms, the third term from the sixth term, 
the fourth term from the second and third terms,
and the fifth term from the seventh and eighth terms.
The form of the solution we consider here is 
\begin{align}
f_0^{\text{sol}}(x,t)&=e^{-qx^2}e^{im_0^{\text{sol}}t} \ , \notag \\
g_0^{\text{sol}}(x,t)&=-e^{-qx^2}e^{im_0^{\text{sol}}t} \ ,
\notag \\
(m_0^{\text{sol}})^2&=-2q 
\left(1-\frac{1}{3}(2\pi\alpha'q)^2
+a(2\pi\alpha'q)^4 \right) +{\cal O} (q^6) \ .\label{f^6sol} 
\end{align}
By substituting the solution (\ref{f^6sol}) into the equations of motion 
(\ref{F^6eom}), we obtain the result 
\begin{align} \notag
\hat{O}_0 \left(
\begin{array}{c}
f_0^{\text{sol}}\\ g_0^{\text{sol}}
\end{array}\right)
&=\left(
\frac{4q}{3}\left((2\pi\alpha'q)^2-3a(2\pi\alpha'q)^4\right)+{\cal O}(q^6)
\right)
\left(
\begin{array}{c}
f_0^{\text{sol}}\\ g_0^{\text{sol}}
\end{array}\right) \ , \\ \notag
\hat{O}_2 \left(
\begin{array}{c}
f_0^{\text{sol}}\\ g_0^{\text{sol}}
\end{array}\right)
&=\left(
-\frac{4q}{3}\big((2\pi\alpha'q)^2+\frac{1}{6}(2\pi\alpha'q)^4\big)
+{\cal O}(q^6)
\right)
\left(
\begin{array}{c}
f_0^{\text{sol}}\\ g_0^{\text{sol}}
\end{array}\right)\ , \\ \notag
\hat{O}_3 \left(
\begin{array}{c}
f_0^{\text{sol}}\\ g_0^{\text{sol}}
\end{array}\right)
&={\cal O}(q^6)
\left(
\begin{array}{c}
f_0^{\text{sol}}\\ g_0^{\text{sol}}
\end{array}\right) \ , \\ \notag
\hat{O}_{4,1} \left(
\begin{array}{c}
f_0^{\text{sol}}\\ g_0^{\text{sol}}
\end{array}\right)
&=\left(
\frac{6q}{5}(2\pi\alpha'q)^4
+{\cal O}(q^6)
\right)
\left(
\begin{array}{c}
f_0^{\text{sol}}\\ g_0^{\text{sol}}
\end{array}\right) \ , \\
\left(\hat{O}_{4,2}+\hat{O}_{4,3} \right)\left(
\begin{array}{c}
f_0^{\text{sol}}\\ g_0^{\text{sol}}
\end{array}\right)
&=\left(
-\frac{8q}{45}(2\pi\alpha'q)^4
+{\cal O}(q^6)
\right)
\left(
\begin{array}{c}
f_0^{\text{sol}}\\ g_0^{\text{sol}}
\end{array}\right) \ ,
\end{align}
and summing up all the equations leads to
\begin{align}
\frac{4q}{3}\big((2\pi\alpha'q)^2-3a(2\pi\alpha'q)^4\big)
&-\frac{4q}{3}\big((2\pi\alpha'q)^2+\frac{1}{6}(2\pi\alpha'q)^4\big)\notag \\
&+
\frac{6q}{5}(2\pi\alpha'q)^4-\frac{8q}{45}(2\pi\alpha'q)^4=0 \ . \label{eoma}
\end{align}
We can easily solve (\ref{eoma}), obtaining the result $a=1/5$.
To obtain the mass eigenvalue for the equation (\ref{F^6eom}),
we diagonalize the time derivative operator through a field
redefinition, which yields
\begin{align} \notag
\left(
\begin{array}{c}
\tilde{f}\\ \tilde{g}
\end{array}\right)
= \left[
\hat{{\bf 1}}
+4(2\pi\alpha')^3 \frac{\zeta (3)}{2 \pi^3} q^2
\left(
\begin{array}{cc}
-\p_t^2+4 q^2 x^2
&-4qx\p_x
\\ 0&-\p_t^2-\p_x^2
\end{array}\right)
-\frac{1}{720}(2\pi\alpha'q)^4 \right. \\
\left.
\left(
\begin{array}{cc}
-\frac{1}{q^2}\p_x^4 +\frac{1}{q^2}\p_x^2 \p_t^2 +4
&\frac{24}{q}\p_x^2+\frac{20}{q}\p_t^2+8qx^2
\\  -\frac{6}{q}x\p_x^3-\frac{6}{ q}\p_x^2+\frac{4}{q}x\p_x\p_t^2+
8qx^3\p_x
& 8x^2 \p_x^2-4x^2\p_t^2-16q^2 x^4-68
\end{array}\right) \right.
\notag \\ \left.
+{\cal O} (q^6)
\right]
\left(
\begin{array}{c}
f\\ g
\end{array}\right) \ .
\end{align}
We obtain the equations of motion for the redefined field as
\begin{align}
\left(\hat{O}_0+\hat{O}_2+{\hat{O}_3}'+{\hat{O}_{4,1}}+{\hat{O}_4}'\right)
\left(
\begin{array}{c}
\tilde{f}\\ \tilde{g}
\end{array}\right)
=0 \ ,
\end{align}
where ${\hat{O}_4}'$ is the operator deformed from 
$\hat{O}_{4,2}+\hat{O}_{4,3}$ through the field redefinition and 
is given as follows:
\begin{align}\notag
{\hat{O}_4}'=&
\frac{1}{180}(2\pi\alpha' q)^4 
\\ \notag
&\left[
16 q^2 x^2\left(
\begin{array}{cc}
-4 +4  x \p_x +x^2
 \p_x^2
&-12qx^2-2qx^3\p_x
\\ 
-2qx^2+2qx^3\p_x
& -4q^2x^4
\end{array}\right) \right. \\
&+8\left(
\begin{array}{cc}
-  \p_x^2-4x\p_x^3-x^2\p_x^4
&2 q(2+13x\p_x+8x^2\p_x^2+x^3\p_x^3 ) \\ 
2q(-1+x\p_x-x^2\p_x^2-x^3\p_x^3)
& q^2(4x^2+16x^3\p_x
+4x^4\p_x^2)
\end{array}\right) \notag \\
&+\frac{1}{q^2}\left(
\begin{array}{cc}
\p_x^6
& -12q \p_x^4 -2qx\p_x^5 \\ 
-2q\p_x^4+2 qx\p_x^5
& 4q^2(4\p_x^2
-4x\p_x^3-x^2\p_x^4)
\end{array}\right)\notag \\ \notag
&-\frac{1}{q}\left(
\begin{array}{cc}
x\p_x^5-13\p_x^4
& 120q\p_x^2
\\ \frac{1}{q}\p_x^6
&-x\p_x^5-18\p_x^4
\end{array}\right) \\ \notag
-\frac{1}{q}&\left(
\begin{array}{cc}
q^2(48 x \p_x
-20x^3\p_x^3-56x^2\p_x^2-48)
& q^3(32x^4\p_x^2+128x^3\p_x-72x^2) \\ 
-12qx^2\p_x^4-48qx\p_x^3 -28q\p_x^2
&q^2(20\p_x^3+124x^2\p_x^2
+88x\p_x-88)
\end{array}\right) \\ 
&-\left. \frac{1}{q}\left( \begin{array}{cc}
32q^4x^5\p_x-16q^4 x^4
& -64q^5x^6 \\
16q^3x^4\p_x^2+64q^3x^3\p_x-24q^3x^2
&-32q^4x^5\p_x-176q^4x^4
\end{array}\right) \right]
+{\cal O}(q^7) \ . 
\end{align}
The time dependence of this solution remains unchanged under the field
redefinition, and therefore we find the mass correction of this mode to be 
$a=1/5$. The corresponding eigen function is obtained as
\begin{align} \notag
\tilde{f}_0^{\text{sol}}&=\left(1-8(2\pi\alpha' q)^3
\frac{\zeta (3)}{2 \pi^3}
(1 +2 q x^2)\right. \notag \\ &\left. \notag
+\frac{1}{180}(2\pi\alpha' q)^4
(1+12qx^2+4q^2x^4)+
{\cal O} (q^6)
\right)e^{-qx^2}e^{im_0^{\text{sol}}t} \ , \\
\tilde{g}_0^{\text{sol}}&=-\left(1-16(2\pi\alpha' q)^3
\frac{\zeta (3)}{2 \pi^3}
qx^2 \notag \right. \\ &\left.
+\frac{1}{90}(2\pi\alpha' q)^4
(10-11qx^2+2q^2x^4)+ {\cal O}(q^6)
\right)e^{-qx^2}e^{im_0^{\text{sol}}t} \ .
\end{align}
Therefore, we obtain the correct string mass spectrum to be
\begin{align} \notag
m_0^2&=-2q\left(1-\frac{1}{3}(2\pi\alpha' q)^3+\frac{1}{5}(2\pi\alpha' q)^5
\right)+{\cal O}(q^6) \\
&=-\frac{\theta}{2\pi\alpha' } +{\cal O}(\theta^6)
\end{align}
by considering the non-abelian Born-Infeld action up to order 
$F^6$ terms.

\section{Discussion}

We have carried out a fluctuation analysis of the non-abelian Born-Infeld 
action proposed in Ref.~\citen{KS}, and 
we obtained the correct string mass spectrum up to order $F^6$ terms.
This is the first nontrivial check of the non-abelian Born-Infeld action
at order $F^6$, and this result indicates that the method 
proposed in Ref.~\citen{FKS} might be applicable to all orders in $F$.
There might exist some clues for determining the non-abelian 
Born-Infeld action to all orders.
In the action at order $F^6$, the terms with covariant derivatives 
cannot be eliminated, and we cannot employ the slowly varying
field approximation to obtain a consistent low energy effective
action. Our analysis might also justify this claim.

In this analysis, we have considered intersecting D-strings 
as a background. 
The fluctuation of the off-diagonal mode around this background
can be analyzed in SU(2) Yang-Mills theory, and the mass spectrum 
of the fluctuation mode is identical up to ${\cal O}(\theta)$
with the well-known worldsheet result. 
We first considered the non-abelian Born-Infeld action with $F^5$ terms,
and we obtained a mass spectrum of the off-diagonal fluctuation mode
that is consistent with that found from the worldsheet analysis
up to ${\cal O}(\theta^3)$.
The eigen functions for the spectrum were also obtained.
Next, we considered the non-abelian Born-Infeld correction for $F^6$ terms,
and the mass for the lowest tachyonic mode has obtained correctly up to
${\cal O} (\theta^5)$. The corresponding eigen function was also obtained.
In the analysis with the $F^6$ terms,
the higher massive mode should be obtained correctly,
but it is difficult to obtain because
a naive ansatz like (\ref{f5ans}) does not satisfy the equations of motion.
Therefore, to find the eigen function for the massive mode,
it is necessary to employ another ansatz.
The correct eigen function might give the correct mass spectrum.
It may be interesting to look for the eigen function and 
check the resulting mass spectrum.
It is also desirable to carry out some other checks for this action.
We considered the case in which there is 
one intersection angle here.
Generalization to a higher dimensional intersecting D-branes system
with two or more intersection angles may also be interesting.


\section*{Acknowledgements}

The author would like to appreciate 
Alexander Sevrin and Alexander Wijns for kind and valuable communications.
While this work was nearly completion, the author became aware of 
Ref.~\citen{SW}
which performs fluctuation analysis of the $F^6$ action
in a system with constant background magnetic field.
The author would like to thank 
Koji Hashimoto, Mitsuhiro Kato and Washington Taylor IV 
for useful discussions and comments.
The author would also like to thank Koji Hashimoto for reading
this manuscript.


\appendix

\section{Calculations of $F^6$ Terms}

Here,
we give the result of the fluctuation analysis term by term.
The quantities $\tilde{L}_{4ij}$ denote the action 
quadratic in the fluctuations 
in the background intersecting brane system,
which is obtained from the action $L_{4ij} (i=0,2,4,j=1,\cdots,8)$.
We carried out this calculation using Mathematica.
We obtain the following results:
\begin{align}\notag
\Str \ \tilde{L}_{401}=
& \frac{1}{5}(\dot{f}^2 +  \dot{g}^2)-
 \Big(4fqg+(f'-2qgx)^2\Big) \ ,
\\ \notag
\Str \ \tilde{L}_{402}=
& -\frac{3}{20}(\dot{f}^2 +  \dot{g}^2)+
\frac{3}{4} \Big(4fqg+(f'-2qgx)^2\Big) \ ,
\\ \notag
\Str \ \tilde{L}_{403}=
& \frac{1}{40}(\dot{f}^2 +  \dot{g}^2)-
\frac{1}{8} \Big(4fqg+(f'-2qgx)^2\Big) \ ,
\end{align}
\begin{align}
\notag
\Str \ \tilde{L}_{421}=&
\frac{2}{9} x^2 (f''-2 qg'x-4qg)^2+\frac{4}{9}q^2 x^2 (f-xf'+2qgx^2)^2 \\
&+\frac{1}{36q^2}(f'''-2qg''x-6qg')^2  \notag
-\frac{1}{18q^2} (\dot{f}''-2q\dot{g}'x)^2
-\frac{2}{9} x^2 (\dot{f}'-2 q\dot{g}x)^2 \\ \notag
&+\frac{8}{9}x^2\dot{f}q\dot{g}
+\frac{1}{36q^2}(\ddot{f}'-2\ddot{g}qx)^2-\frac{1}{9q}\ddot{f}\ddot{g} \ ,
\\ \notag
\Str \ \tilde{L}_{422}=&
-\frac{1}{9} x^2 (f''-2 qg'x-4qg)^2-\frac{2}{9} q^2x^2 (f-xf'+2qgx^2)^2
 \\ \notag
&-\frac{1}{72q^2}(f'''-2qg''x-6qg')^2 
+\frac{1}{36q^2} (\dot{f}''-2q\dot{g}'x)^2
+\frac{1}{9} x^2 (\dot{f}'-2 q\dot{g}x)^2 \\ \notag
&-\frac{4}{9}x^2\dot{f}q\dot{g}
-\frac{1}{72q^2}(\ddot{f}'-2\ddot{g}qx)^2+\frac{1}{18q}\ddot{f}\ddot{g}
 \ ,
\\ \notag
\Str \ \tilde{L}_{423}=
& \frac{1}{18}(\dot{f}^2+\dot{g}^2)+\frac{1}{9}\Big(4fqg+(f'-2qgx)^2\Big)
+\frac{2}{9}\dot{f}q\dot{g}x^2 +\frac{1}{18q}\dot{f'}\dot{g'}
\ , \\ \notag 
\Str \ \tilde{L}_{424}=
& -\frac{1}{3}(\dot{f}^2+\dot{g}^2)
\ , \\ \notag  
\Str \ \tilde{L}_{425}=
& 0
\ , \\ \notag 
\Str \ \tilde{L}_{426}=
& -\frac{1}{6}(\dot{f}^2+\dot{g}^2)
+\frac{2}{9}\dot{f}q\dot{g}x^2
+\frac{1}{18q}\dot{f}'\dot{g}'
\ , \\ \notag 
\Str \ \tilde{L}_{427}=
& \frac{2}{9}(\dot{f}^2+\dot{g}^2)
-\frac{8}{9}\dot{f}q\dot{g}x^2-\frac{2}{9q}\dot{f'}\dot{g}'
\ , \\ \notag 
\Str \ \tilde{L}_{428}=
&0 \ ,
\end{align}
\begin{align}
\notag 
\Str \ \tilde{L}_{441}=&
\frac{4}{45} x^2 (f''-2 qg'x-4qg)^2
+\frac{8}{45} q^2x^2 (f-xf'+2qgx^2)^2 \\ \notag
&+\frac{1}{90q^2}(f'''-2qg''x-6qg')^2 
-\frac{4}{45}(\dot{f}^2+\dot{g}^2)
-\frac{1}{45q^2} (\dot{f}''-2q\dot{g}'x)^2 \\ \notag
&-\frac{4}{45} x^2 (\dot{f}'-2 q\dot{g}x)^2
-\frac{8}{45}x^2\dot{f}q\dot{g}
-\frac{6}{45q}\dot{f}'\dot{g}'
+\frac{1}{90q^2}(\ddot{f}'-2\ddot{g}qx)^2 \notag \ , 
\end{align}
\begin{align} \notag 
\Str \ \tilde{L}_{442}= &0 \ , 
\\ \notag 
\Str \ \tilde{L}_{443}=
& -\frac{1}{15}(\dot{f}^2+\dot{g}^2)
-\frac{4}{45}\dot{f}q\dot{g}x^2
-\frac{1}{45q}\dot{f'}\dot{g'} \ ,
\\ \notag 
\Str \ \tilde{L}_{444}=
& -\frac{1}{5}(\dot{f}^2+\dot{g}^2)
-\frac{4}{15}\dot{f}q\dot{g}x^2
-\frac{1}{15q}\dot{f'}\dot{g'} \ ,
\\ \notag 
\Str \ \tilde{L}_{445}=
& \frac{16}{45}\dot{f}q\dot{g}x^2
+\frac{4}{45q}\dot{f'}\dot{g}' \ ,
\\ \notag 
\Str \ \tilde{L}_{446}=
& \frac{4}{45}(\dot{f}^2+\dot{g}^2) \ ,
\\ \notag 
\Str \ \tilde{L}_{447}=
&
 \frac{2}{45}(\dot{f}^2+\dot{g}^2)+\frac{4}{45}\Big(4fqg+(f'-2qgx)^2\Big) \ ,
\\ \notag
&+\frac{8}{45}\dot{f}q\dot{g}x^2+\frac{2}{45q}\dot{f}'\dot{g}'
\\ \notag 
\Str \ \tilde{L}_{448}= 
&0 \ .
\end{align}
Summing up all the fluctuations,
we obtain (\ref{L4-1}) and (\ref{L4-23}).

\newcommand{\J}[4]{{\sl #1} {\bf #2} (#3) #4}


\begin{thebibliography}{99}

\bibitem{wit}
E.~Witten,
\NPB{460,1996,335}; hep-th/9510135.

\bibitem{tse}
A.~A.~Tseytlin,
\NPB{276,1986,391}; [Errata {\bf 291} (1987), 876];
\NPB{501,1997,41}; hep-th/9701125.

\bibitem{BDL}
M.~Berkooz, M.~R.~Douglas and R.~G.~Leigh, 
\NPB{480,1996,265}; hep-th/9606139.

\bibitem{AkiWati}
A.~Hashimoto and W.~Taylor IV,
\NPB{503,1997,193}: hep-th/9703217.

\bibitem{DST}
F.~Denef, A.~Sevrin and J.~Troost,
\NPB{581,2000,135}; hep-th/0002180 ;

A.~Sevrin, J.~Troost and W.~Troost,
\NPB{603,2001,389}; hep-th/0101192.

\bibitem{FKS}
Lies De Fosse, Paul Koerber and Alexander Sevrin,
\NPB{603,2001,413}; hep-th/0103015.

\bibitem{F^5action}
P.~Koerber and A.~Sevrin, 
\JHEP{10,2001,003}; hep-th/0108169.

\bibitem{F^5fluc}
P.~Koerber and A.~Sevrin,
\JHEP{09,2001,009};
hep-th/0109030.

\bibitem{MBM}
R.~Medina, F.~T.~Brandt and F.~R.~Machado,
\JHEP{07,2002,071}, hep-th/0208121.

\bibitem{KS}
P.~Koerber and A.~Sevrin,
\JHEP{10,2002,046}; hep-th/0208044.

\bibitem{Recombi}
K.~Hashimoto and S.~Nagaoka,
\JHEP{06,2003,034}; hep-th/0303204.

\bibitem{Morosov}
A.~V.~Morosov,
\PLB{433,1998,291}; hep-th/9803110.

\bibitem{Bilal}
A.~Bilal, 
\NPB{618,2001,21}; {\tt hep-th/0106062}.

\bibitem{SW}
A.~Sevrin and A.~Wijns, hep-th/0306260. 


\end{thebibliography}
\end{document}